%% file: edgespeechnets.tex
\documentclass{article}

\PassOptionsToPackage{numbers, compress}{natbib}
 \usepackage[final]{nips_2018}

\usepackage[utf8]{inputenc} % allow utf-8 input
\usepackage[T1]{fontenc}    % use 8-bit T1 fonts
\usepackage{hyperref}       % hyperlinks
\usepackage{url}            % simple URL typesetting
\usepackage{booktabs}       % professional-quality tables
\usepackage{amsfonts}       % blackboard math symbols
\usepackage{nicefrac}       % compact symbols for 1/2, etc.
\usepackage{microtype}      % microtypography
\usepackage[cmex10]{amsmath}
\usepackage{amssymb}
\usepackage{graphicx}

\title{EdgeSpeechNets: Highly Efficient Deep Neural Networks for Speech Recognition on the Edge}

\author{
  Zhong Qiu Lin$^{*,1,3}$, Audrey G. Chung$^{1,2,3}$, Alexander Wong$^{1,2,3}$\\
  $^{1}$ Vision and Image Processing Research Group, University of Waterloo, Waterloo, ON, Canada\\
  $^{2}$ Waterloo Artificial Intelligence Institute, University of Waterloo, Waterloo, ON, Canada\\
  $^{3}$ DarwinAI Corp., Waterloo, ON, Canada \\
  \texttt{$^{*}$ zq2lin@edu.uwaterloo.ca} \\
}

\begin{document}

\maketitle

\begin{abstract}
  \input{./content/Abstract.tex}
\end{abstract}
\vspace{-0.15in}
\section{Introduction}
\vspace{-0.15in}
\label{Introduction}
\input{./content/Introduction.tex}

\section{Methods}
\vspace{-0.15in}
\label{Methods}
\input{./content/Methods.tex}
\vspace{-0.25in}
\section{Results and Discussion}
\vspace{-0.15in}
\label{Results}
\input{./content/Results.tex}

\vspace{-0.13in}
\subsubsection*{Acknowledgements}
\vspace{-0.1in}
This work was supported by NSERC, Canada Research Chairs Program, and DarwinAI Corp.
\vspace{-0.15in}
\footnotesize
\bibliographystyle{IEEEtran}
\bibliography{edgespeechnets}

\end{document}

%% file: content/Abstract.tex
Despite showing state-of-the-art performance, deep learning for speech recognition remains challenging to deploy in on-device edge scenarios such as mobile and other consumer devices. Recently, there have been greater efforts in the design of small, low-footprint deep neural networks (DNNs) that are more appropriate for edge devices, with much of the focus on design principles for hand-crafting efficient network architectures. In this study, we explore a human-machine collaborative design strategy for building low-footprint DNN architectures for speech recognition through a marriage of human-driven principled network design prototyping and machine-driven design exploration. The efficacy of this design strategy is demonstrated through the design of a family of highly-efficient DNNs (nicknamed \textbf{EdgeSpeechNets}) for limited-vocabulary speech recognition. Experimental results using the Google Speech Commands dataset for limited-vocabulary speech recognition showed that EdgeSpeechNets have higher accuracies than state-of-the-art DNNs (with the best EdgeSpeechNet achieving $\sim$97\% accuracy), while achieving significantly smaller network sizes (as much as $7.8\times$ smaller) and lower computational cost (as much as $36\times$ fewer multiply-add operations, $10\times$ lower prediction latency, and $16\times$ smaller memory footprint on a Motorola Moto E phone), making them very well-suited for on-device edge voice interface applications.

%% file: content/Introduction.tex
Deep learning has seen widespread interest in recent years, and has been demonstrated to achieve state-of-the-art performance for a wide range of applications in speech recognition.  In particular, limited-vocabulary speech recognition~\cite{Warden2018}, also known as keyword spotting, has recently seen significant interest as an important application of deep learning for mobile, IoT, and other edge devices.  The ability to rapidly recognize specific keywords from a stream of verbal utterances can enable voice interfaces with which the user can interact in a natural, verbal manner without the need for cloud computing, which is particularly important in scenarios where privacy and internet connectivity are of concern.

Despite the promises, deep learning for speech recognition tasks such as limited-vocabulary speech recognition remains challenging to deploy in on-device edge scenarios such as mobile and other consumer devices due to computational and memory requirements. As such, there has been greater recent efforts to design small, low-footprint deep neural network (DNN) architectures that are more appropriate for edge devices, with much of the focus on design principles for hand-crafting efficient network architectures~\cite{Warden2018,Sainath2015,Tang2017}.  In this study, we explore a human-machine collaborative design strategy for building low-footprint DNN architectures for speech recognition through a marriage of human-driven principled network design prototyping and machine-driven design exploration via generative synthesis~\cite{Wong2018}.  More specifically, a family of highly-efficient DNNs (nicknamed \textbf{EdgeSpeechNets}) are designed for limited-vocabulary speech recognition using this strategy.

%% file: content/Methods.tex
The human-machine collaborative design strategy presented in this study for building \textbf{EdgeSpeechNets} (low-footprint DNN architectures for limited-vocabulary speech recognition) comprises of two main steps.  First, design principles are leveraged to construct an initial design prototype catered towards the task of limited-vocabulary speech recognition.  Second, machine-driven design exploration is performed based on the constructed initial design prototype and a set of design requirements to generate a set of alternative highly-efficient DNN designs appropriate to the problem space.  As such, the goal of this strategy is to combine human ingenuity with the meticulousness of a machine. Details of these two steps are described below.

\subsection{Human-driven Design Prototyping}
\vspace{-0.1in}
The first step of the presented design strategy is to leverage design principles to construct an initial design prototype catered towards the task of limited-vocabulary speech recognition.  Based on past literature, a very effective strategy for leveraging deep learning for limited-vocabulary speech recognition is to first transform the input audio signal into mel-frequency cepstrum coefficient (MFCC) representations.  Inspired by~\cite{Tang2017_Honk}, the input layer of the design prototype takes in a two-dimensional stack of MFCC representations using a 30ms window with a 10ms time shift across a one-second band-pass filtered (cutoff from 20Hz to 4kHz for reducing noise) audio sample.

\begin{figure}
%\vspace{- 0.6 cm}
\centering
	\includegraphics[width = \linewidth]{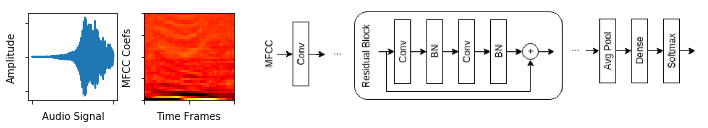}
	\caption{(a) Audio signal; (b) MFCC representations; (c) initial design prototype.}	
	\label{fig:prototype}
\vspace{-0.2 cm}
\end{figure}

For the intermediate representation layers of the initial design prototype, we leverage the concept of deep residual learning~\cite{He2016} and specify the use of residual blocks comprised of alternating convolution and batch normalization layers, with skip connections between residual blocks.  Networks built around deep residual stacks have been shown to enable easier learning of DNNs with greater representation capabilities, and has been previously demonstrated to enable state-of-the-art speech recognition performance~\cite{Tang2017}. After the intermediate representation layers, average pooling is specified in the initial design prototype, followed by a dense layer. Finally, a softmax layer is defined as the output of the initial design prototype to indicate which of the keywords was detected from the verbal utterance.  Driven by these design principles, the initial design prototype is shown in Figure~\ref{fig:prototype}.

\subsection{Machine-driven Design Exploration}
\vspace{-0.1in}
The second step in the presented design strategy is to perform machine-driven design exploration based on the initial design prototype, while considering a set of design requirements that ensure the generation of a set of alternative highly-efficient DNN designs appropriate for on-device limited-vocabulary speech recognition. Previous literature~\cite{Tang2017} has leveraged manual design exploration for designing low-footprint DNNs for speech recognition, where high-level parameters such as the width and depth of a network are varied to observe the trade-offs between resource usage and accuracy in a similar vein as~\cite{Howard2017,Sandler2018}.  However, the coarse-grained nature of such a design exploration strategy can be quite limiting in terms of the diversity in the network architectures that can be found.  Motivated to overcome these limitations, we instead take advantage of a highly flexible machine-driven design exploration strategy in the form of generative synthesis~\cite{Wong2018}.  Briefly, the goal of generative synthesis is to learn a generator $\mathcal{G}$ that, given a set of seeds $S$, can generate DNNs $\left\{N_s|s \in S\right\}$ that maximize a universal performance function $\mathcal{U}$ (e.g.,~\cite{Wong2018_Netscore}) while satisfying requirements defined by an indicator function $1_r(\cdot)$, i.e.,
\vspace{-0.05in}
\begin{equation}
\mathcal{G}  = \max_{\mathcal{G}}~\mathcal{U}(\mathcal{G}(s))~~\textrm{subject~to}~~1_r(\mathcal{G}(s))=1,~~\forall s \in S.
\label{optimization}
%\vspace{-0.17in}
\end{equation}
An approximate solution to this optimization problem is found in an progressive manner, with the generator initialized based on prototype $\varphi$, $\mathcal{U}$, and $1_r(\cdot)$ and a number of successive generators being constructed.  We take full advantage of this interesting phenomenon by leveraging this set of generators to synthesize a family of EdgeSpeechNets that satisfies these requirements.  More specifically, we configure the indicator function $1_r(\cdot)$ such that the validation accuracy $\geq$ 95\% on the Google Speech Commands dataset~\cite{Warden2017}.

\subsection{Final Architecture Designs}
\vspace{-0.1in}
The architecture design of the EdgeSpeechNets produced using the presented human-machine collaborative strategy are summarized in Tables~\ref{tab_EdgeSpeech1} and~\ref{tab_EdgeSpeech3}. It can be observed that the architecture designs produced have diverse architectural differences that can only be achieved via fine-grained machine-driven design exploration.
{\scriptsize
\begin{table}[h]
	\caption{Network architectures of EdgeSpeechNet-A (left) and EdgeSpeechNet-B (right)}
\scriptsize
\center
\begin{tabular}{r|ccc|cc}
	\hline
	\textbf{Type}	&	$m$	&	$r$	&	$n$	&	\textbf{Params}\\
	\hline
	conv			&	3	&	3	&	39	&	351	\\
	conv	&	3	&	3	&	20	&	7020	\\
	conv 	&	3	&	3	&	39	&	7020	\\
	conv	&	3	&	3	&	15	&	5265	\\
	conv 	&	3	&	3	&	39	&	5265	\\
	conv	&	3	&	3	&	25	&	8775	\\
	conv 	&	3	&	3	&	39	&	8775\\
	conv	&	3	&	3	&	22	&	7722	\\
	conv 	&	3	&	3	&	39	&	7722\\
	conv	&	3	&	3	&	22	&	7722\\
	conv 	&	3	&	3	&	39	&	7722\\
	conv	&	3	&	3	&	25	&	8775\\
	conv 	&	3	&	3	&	39	&	8775\\
	conv			&	3	&	3	&	45	&	15795\\
	avg-pool		&	-	&	-	&	-	&	-	\\
	dense	&	-	&	-	&	12	&	540	\\
	softmax			&	-	&	-	&	-	&	-	\\
	\hline \hline
	Total			&	-	&	-	&	-	&	107K
	\label{tab_EdgeSpeech1}
	\end{tabular}
\quad
\scriptsize
	\begin{tabular}{r|ccc|cc}
	\hline
	\textbf{Type}	&	$m$	&	$r$	&	$n$	&	\textbf{Params}	\\
	\hline
	conv			&	3	&	3	&	30	&	270\\
	conv	&	3	&	3	&	8	&	2160	\\
	conv 	&	3	&	3	&	30	&	2160	\\
	conv	&	3	&	3	&	9	&	2430	\\
	conv 	&	3	&	3	&	30	&	2430	\\
	conv	&	3	&	3	&	11	&	2970	\\
	conv 	&	3	&	3	&	30	&	2970	\\
	conv	&	3	&	3	&	10	&	2700	\\
	conv 	&	3	&	3	&	30	&	2700	\\
	conv	&	3	&	3	&	8	&	2160	\\
	conv 	&	3	&	3	&	30	&	2160	\\
	conv	&	3	&	3	&	11	&	2970	\\
	conv 	&	3	&	3	&	30	&	2970	\\
	conv			&	3	&	3	&	45	&	12150	\\
	avg-pool		&	-	&	-	&	-	&	-	\\
	dense		&	-	&	-	&	12	&	540	\\
	softmax			&	-	&	-	&	-	&	-	\\
	\hline \hline
	Total			&	-	&	-	&	-	&	43.7K
	\label{tab_EdgeSpeech2}
	\end{tabular}
	\caption{Network architectures of EdgeSpeechNet-C (left) and EdgeSpeechNet-D (right)}
\scriptsize
	\begin{tabular}{r|ccc|cc}
	\hline
	\textbf{Type}	&	$m$	&	$r$	&	$n$	&	\textbf{Params}	\\
	\hline
	conv			&	3	&	3	&	24	&	216	\\
	conv	&	3	&	3	&	6	&	1296	\\
	conv 	&	3	&	3	&	24	&	1296	\\
	conv	&	3	&	3	&	9	&	1944	\\
	conv 	&	3	&	3	&	24	&	1944	\\
	conv	&	3	&	3	&	12	&	2592	\\
	conv 	&	3	&	3	&	24	&	2592	\\
	conv	&	3	&	3	&	6	&	1296	\\
	conv 	&	3	&	3	&	24	&	1296	\\
	conv	&	3	&	3	&	5	&	1080	\\
	conv 	&	3	&	3	&	24	&	1080	\\
	conv	&	3	&	3	&	6	&	1296	\\
	conv 	&	3	&	3	&	24	&	1296	\\
	conv	&	3	&	3	&	2	&	432	\\
	conv 	&	3	&	3	&	24	&	432	\\
	conv			&	3	&	3	&	45	&	9720	\\
	avg-pool		&	-	&	-	&	-	&	-	\\
	dense		&	-	&	-	&	12	&	540	\\
	softmax			&	-	&	-	&	-	&	-	\\
	\hline \hline
	Total			&	-	&	-	&	-	&	30.3K
	\label{tab_EdgeSpeech4}
	\end{tabular}
\quad
\scriptsize
\begin{tabular}{r|ccc|cc}
	\hline
	\textbf{Type}	&	$m$	&	$r$	&	$n$	&	\textbf{Params}	\\
	\hline
	conv			&	3	&	3	&	45	&	405\\
	avg-pool		&	-	&	-	&	-	&	-	\\
	conv	&	3	&	3	&	30	&	12150	\\
	conv 	&	3	&	3	&	45	&	12150	\\
	conv	&	3	&	3	&	33	&	13365	\\
	conv 	&	3	&	3	&	45	&	13365	\\
	conv	&	3	&	3	&	35	&	14175	\\
	conv 	&	3	&	3	&	45	&	14175	\\
	avg-pool		&	-	&	-	&	-	&	-	\\
	dense		&	-	&	-	&	12	&	540	\\
	softmax			&	-	&	-	&	-	&	-	\\
	\hline \hline
	Total			&	-	&	-	&	-	&	80.3k
	\label{tab_EdgeSpeech3}
	\end{tabular}
\end{table}	
} 

%% file: content/Results.tex
\begin{table}[h]
	\centering
	\caption{Test accuracy of EdgeSpeechNets in comparison to trad-fpool13~\cite{Sainath2015}, tpool2~\cite{Sainath2015}, res15~\cite{Tang2017}, and res15-narrow~\cite{Tang2017}, NetScores, and model sizes in terms of number of parameters and multiply-add operations. All results are the mean across 5 runs.  Best results are in \textbf{bold}.}
	\begin{tabular}{p{3cm}cc|cc}
		\hline
		\textbf{Model}& \textbf{Test Accuracy} & \textbf{NetScore} & \textbf{Params} &	\textbf{Mult-Adds} \\
		\hline %\hline
		trad-fpool13~\cite{Sainath2015}	&	$90.5\%$	&	$85.93$	&	1.37M	&	125M \\
		tpool2~\cite{Sainath2015}		&	$91.7\%$	&	$87.99$	&	1.09M	&	103M \\
		\hline
		res15~\cite{Tang2017} 			&	$95.8\%$	&	$85.98$	&	238K 	&	894M \\
		res15-narrow~\cite{Tang2017} 	&	$94.0\%$ 	&	$100.59$	&	42.6K 	&	160M \\
		\hline
		EdgeSpeechNet-A		& 	\textbf{96.8$\%$}	&	$93.75$		&	107K	&	343M \\
		EdgeSpeechNet-B		& 	$96.3\%$	&	$101.82$	&	43.7K	&	126M \\
		EdgeSpeechNet-C		& 	$96.2\%$	&	$105.12$	&	\textbf{30.3K}	&	83.5M \\
		EdgeSpeechNet-D		& 	$95.8\%$	&	\textbf{106.67}	&	80.3K	&	\textbf{24.5M} \\
		\hline
	\end{tabular}
	\label{tab_Results}
\vspace{-0.15in}
\end{table}	

The efficacy of the produced EdgeSpeechNets were evaluated using the Google Speech Commands dataset~\cite{Warden2017}~\footnote{https://research.googleblog.com/2017/08/
launching-speech-commands-dataset.html}. The Speech Commands Dataset was designed for limited-vocabulary speech recognition and contains 65,000 one-second samples of 30 short words and background noise samples.  For comparison purposes, the results for two state-of-the-art deep neural networks presented in~\cite{Tang2017} (res15 and res15-narrow) and the Google networks presented in~\cite{Sainath2015} (trad-fpool13, tpool2) were also presented.  As shown in Table~\ref{tab_Results}, the produced EdgeSpeechNets had higher accuracies at much smaller sizes and lower computational costs than state-of-the-art deep neural networks.  In terms of best accuracy, EdgeSpeechNet-A achieved \textbf{1\%} higher accuracy compared to the state-of-the-art res15~\cite{Tang2017} while having \textbf{>2.2$\times$} fewer parameters and requiring \textbf{>2.6$\times$} fewer multiply-add operations.  In fact, the best of 5 runs for EdgeSpeechNet-A achieved a test accuracy reaching \textbf{$\sim$97\%}, thus noticeably outperforming previously published results.  More interesting, EdgeSpeechNet-B still achieved higher accuracy (\textbf{0.5\%} higher) compared to res15 while having \textbf{>5.4$\times$} fewer parameters and requiring \textbf{$\sim$7.1$\times$} fewer multiply-add operations.  In terms of smallest size, EdgeSpeechNet-C achieved higher accuracy (\textbf{0.4\%} higher) compared to res15 but has \textbf{>7.8$\times$} fewer parameters and requiring \textbf{>10.7$\times$} fewer multiply-add operations.  In terms of lowest computational cost, EdgeSpeechNet-D achieved the same accuracy compared to res15 but requires \textbf{$\sim$36.5$\times$} fewer multiply-add operations.  When compared to the Google network tpool2~\cite{Sainath2015}, EdgeSpeechNet-D achieved \textbf{4.1\%} higher accuracy while having \textbf{>13.5$\times$} fewer parameters and requiring \textbf{$\sim$4.2$\times$} fewer multiply-add operations.  In terms of the highest NetScore, EdgeSpeechNet-D achieved a NetScore that is >\textbf{20} points than res15, which demonstrates a strong balance between accuracy, computational cost, and size.  Finally, running on a 1.4 GHz Cortex-A53 mobile processor in a Motorola Moto E phone using TensorFlow Mobile, EdgeSpeechNet-D ran with an average prediction latency of 34ms and memory footprint of $\sim$1MB (\textbf{>10$\times$} lower latency and \textbf{>16.5$\times$} smaller memory footprint than res15).  These results demonstrate that the EdgeSpeechNets were able to achieve state-of-the-art performance while still being noticeably smaller and requiring significantly fewer computations, making them very well-suited for on-device edge voice interface applications.  Given the promising prospects of the presented human-machine collaborative design strategy, we aim to further explore this strategy for designing highly-efficient deep neural networks in other applications such as visual perception and natural language processing.